# Extracting predictive information from heterogeneous data streams using Gaussian Processes

S. Ghoshal[a,b,*] and S. Roberts[a,b,1]
[a]*Department of Engineering Science, University of Oxford, UK*
[b]*Oxford-Man Institute of Quantitative Finance, University of Oxford, UK*

**Abstract**. Financial markets are notoriously complex environments, presenting vast amounts of noisy, yet potentially informative data. We consider the problem of forecasting financial time series from a wide range of information sources using online Gaussian Processes with Automatic Relevance Determination (ARD) kernels. We measure the performance gain, quantified in terms of Normalised Root Mean Square Error (NRMSE), Median Absolute Deviation (MAD) and Pearson correlation, from fusing each of four separate data domains: time series technicals, sentiment analysis, options market data and broker recommendations. We show evidence that ARD kernels produce meaningful feature rankings that help retain salient inputs and reduce input dimensionality, providing a framework for sifting through financial complexity. We measure the performance gain from fusing each domain's heterogeneous data streams into a single probabilistic model. In particular our findings highlight the critical value of options data in mapping out the curvature of price space and inspire an intuitive, novel direction for research in financial prediction.

## 1. Introduction

One of the central challenges in financial forecasting is determining where to look. A financial instrument's time series history, comparables and derivatives, news articles and opinion pieces all have the potential to influence price evolution. Developing a robust framework for knowledge extraction from disparate, jointly informative datasets remains an open challenge for the finance and machine learning communities.

In this paper we forecast daily returns on the S&P500 index, a broad market benchmark for US equities commonly viewed as a gauge of financial stability. The S&P500 is a market capitalisation-weighted index of the 500 largest corporations in the US, covering the full range of technology, consumer goods, utilities and financial services companies. It is one of the most visible benchmarks in the world, actively traded by buy-and-hold mutual funds and high-frequency hedge funds alike.

We begin by postulating four broad categories in which to search for salient explanatory variables. *Market technicals* include lagged returns to measure autocorrelation, as well as chartist signals used in industry like the Moving Average Convergence Divergence (MACD). *Sentiment analysis* covers the impact of newsflow, measured by optimism or pessimism in social media. *Options market metrics* provide a glimpse into the positioning of market experts and give us a principled, data-driven method for modelling price space as an inhomogeneous dimension with regions of directional bias and return compression. *Broker recommendations* collate the wisdom of equity analysts and allow us to measure the predictive value, if any, of their upgrades and downgrades.

---

[1] E-mail: sjrob@robots.ox.ac.uk.

[*]Corresponding author: S. Ghoshal. Tel.: +44 7740 700665; E-mail: sghoshal@robots.ox.ac.uk.

We show that predictive performance improves when combining signals from each domain, and provide a principled framework for the triage of inputs by implementing Automatic Relevance Determination (ARD) in the covariance parametrisation of an adaptive Gaussian Process model. The ranking that emerges from this analysis defies expectations, and encourages further investigation of options markets and the price space representation.

## 2. Prior work

We first establish context for our study by reviewing relevant prior research in the domain of financial prediction using each of our various data streams. We then turn our attention to common practice multivariate analysis techniques.

Technical analysis was one of the earliest forms of financial forecasting, first appearing in merchant accounts of the Dutch markets in the 17th century. Formalised as a discipline in the 1940s (Edwards and Magee, 1946), it involves the use of price and volume time series to make directional forecasts. It has been extensively studied in prior regression analyses (Lo et al., 2000) demonstrating the incremental gains in predictive performance provided by identifying specific patterns in price history. Technicals-driven Gaussian Process regression has been applied to forecasting time-series in a wide range of asset classes, including stock market prices (Farrell and Correa, 2007), stock market volatility (Ou and Wang, 2009) and commodity spreads (Chapados and Bengio, 2007). These studies show that model performance is highly reliant on the size of the training set.

Literature on financial prediction using text data has proliferated in recent decades, closely tracking advances in the field of natural language processing. The methodology in this domain has typically involved converting words or phrases into numerical gauges of sentiment with which to predict stock market direction (Nikfarjam et al., 2010). Modelling techniques have ranged from simple Naive-Bayes or Support Vector Machine classifiers to more advanced algorithms built on deep learning. More recent work in sentiment composition has sought to predict economic indicators like the U.S. Non-Farm Payrolls using newsflow data. These studies show evidence that accurate parsing of news articles can produce state-of-the-art forecasts for market-moving announcements (Levenberg et al., 2013, 2014).

Research on the interactions between stock and options market prices has been scarce, though early attempts were made to assess correlation in the volume data. Studies indicated that call options flow lead underlying shares flow with a one-day lag, lending credence to the hypothesis of a sequential flow of information between the options and stock markets (Anthony, 1998).

Multiple studies have been conducted to ascertain the influence of buy and sell recommendations on stock prices. Research on equity analyst reports show significant, systematic but asymmetric drift in the aftermath of broker actions, with short-lived, modest gains following upgrades but durable, material sell-offs following downgrades (Womack, 1996). The magnitude of these changes depended not only on the action (upgrade vs. downgrade), but also on the reputation of the analyst, the size of their brokerage firm and the size of the recommended firm (Stickel, 1995).

Various techniques have been applied to multivariate analysis in finance, relying on Independent Component Analysis to reduce dimensionality (Lu et al., 2009) and elliptical copula models to capture input dependencies (Biller and Corlu, 2012). These studies find incremental information gain in using multiple time series from the same domain. By contrast, our work focuses on heterogeneous data fusion and modelling inter-domain dependencies.

## 3. Data

In this section we detail the features considered for each of the four domains under consideration, all of which will be used to predict *Return(t+1)*, the next-day log-returns on the S&P500.

### 3.1. Technical indicators

Market technicals are metrics derived directly from the price history p(t) of a financial instrument. We consider four features commonly watched in industry (Taylor and Allen, 1992): the *previous daily log-return* on the S&P500, its *50-day Simple Moving Average*, as well as the *Moving Average Convergence Divergence (MACD)* and *Signal Line*, constructed from Exponential Moving Averages (EMA) of the time series as follows:

$$\text{MACD(t)} = \text{12-day EMA}\bigl[p(t)\bigr] - \text{26-day EMA}\bigl[p(t)\bigr] \tag{1}$$

$$\text{Signal Line(t)} = \text{MACD(t)} - 9\text{-day EMA}\big[\text{MACD(t)}\big] \tag{2}$$

We do not believe that the formulation of these metrics is inherently meaningful, but rather that standardised definitions provide precise, measurable thresholds at which chartist market participants will react. Including these features will allow our model to identify those thresholds and thereby anticipate technically-led order flow.

### 3.2. Sentiment analysis

While factual newsflow is significant, it is specifically the polarity of its interpretation by markets - as beats or disappointments - that drives market movement. Market sentiment was captured using indicators derived from both Twitter and Stocktwits, a social media site dedicated to real-time discussions of financial markets and actively frequented by S&P500 retail investors. Two further metrics were derived by tracking the daily changes in the sentiment indices.

### 3.3. Options-based modelling of price space

As a province reserved for more sophisticated traders, options market open interest volumes offer a window into the expectations of the most experienced, well-capitalised participants. As strike-sensitive instruments[1], options data also allow us to gauge how these expectations vary at different price levels, motivating the representation of price as an inhomogeneous space with identifiable regions of high directional bias or variance. Illustratively, high open interest (OI) in call options coupled with low open interest in put options indicates experts pre-positioning for a rally. By contrast, high open interest in straddles[2] at a given strike implies low consensus among experts about *directionality* at that price, and hints at evenly matched, competing forces that will compress returns locally. We term this phenomenon *viscosity*, appealing to the visual analogy of price space as an inhomogeneous fluid that enables price gaps in regions of low viscosity and prevents it in regions of high viscosity.

---

[1]Call and put option prices are calculated via the Black-Scholes formula and depend on a 'strike' level that defines the price at which the option owner may buy or sell the underlying asset.

[2]A long straddle position refers to the ownership of both a call and a put option at the same strike price and expiry date: it does not express a directional view, and benefits so long as the underlying asset deviates sufficiently from the strike before expiry.

To capture the directionality and viscosity implied by open interest data, we constructed two metrics. Directionality measures the daily change in call minus put open interest at strike *s* with time-to-expiry $\tau$, summed across all strikes *S* and expiries *T*. It proxies for expert optimism as evidenced by bullish option positioning, and by construction correlates positively with S&P500 next-day returns. The scaling factors $e^{-\gamma_D \tau}$ account for the time sensitivity of options traders, and serve to scale up the weight of nearby expiries by mimicking the exponential decay of gamma risk as time-to-expiry lengthens.

$$\begin{aligned}\text{Directionality(t)} = &\sum_{s \in S, \tau \in T} \Big[(\text{OI(s,t,}\tau)_{Call} \\ &- \text{OI(s,t,}\tau)_{Put}) \times \exp(-\gamma_D \tau)\Big] \\ &- \sum_{s \in S, \tau \in T} \Big[(\text{OI(s,t-1,}\tau)_{Call} - \text{OI(s,t-1,}\tau)_{Put}) \\ &\times \exp(-\gamma_D \tau)\Big] \end{aligned} \tag{3}$$

The parameter $\gamma_D$ measures the rate at which Directionality decays as a function of time-to-expiry, and is optimised over the training data by solving:

$$\gamma_D = \arg\max_{\gamma_D} \text{corr}\big(\text{Directionality, Return}\big) \tag{4}$$

where corr is the linear correlation between its two arguments.

In parametrising viscosity, we make three modelling assumptions. Firstly, the pinning effect of high straddle open interest is at its greatest for options very near their expiry date. Secondly, this effect decays as live prices move away from the straddle's strike *s*. Thirdly, we claim that open interest volumes follow a lognormal distribution, evolving over time through the compounding of normally-distributed exponential factors and restricted to non-negative values. These claims jointly motivate the following representation:

$$\begin{aligned}\text{Viscosity(t)} = &\sum_{s \in S, \tau \in T} \\ &\Big[\exp(-\lambda_V |\text{price(t)-s}|) \times \exp(-\gamma_V \tau) \\ &\times \log[\min(\text{OI(s,t,}\tau)_{Call}, \text{OI(s,t,}\tau)_{Put}) + 1]\Big] \end{aligned} \tag{5}$$

We expect a significant negative correlation between viscosity and the magnitude of S&P500 next-day returns; as such tuning $\lambda_V$ and $\gamma_V$ equates to solving the following optimisation problem:

$$\lambda_V, \gamma_V = \arg \min_{\lambda_V, \gamma_V} \text{corr}\big(\text{Viscosity}, |\text{Return}|\big) \quad (6)$$

### 3.4. Broker recommendations

Market analysts issue recommendations on individual stocks rather than on the broad market - partly a reflection of the incentive structure for brokerage firms: commissions are substantially larger for actively managed portfolios than for passive index-trackers.

To overcome this, we construct an index of broker opinions, based on a weighted sum of broker recommendations across the top 100 stocks in the S&P500. These account for 63% of the index's market capitalisation, and broker actions on these household names have a disproportionate effect on the index as a whole. Two indices were built from these weighted sums, to track both changes in analyst opinion (upgrades and downgrades) and the consensus state (buy, hold or sell).

## 4. ARD Gaussian Processes

We briefly recall the fundamentals of Gaussian Process modelling before describing ARD kernels and the associated notion of relevance score. For a comprehensive treatment of Gaussian Processes, please refer to Rasmussen and Williams (2006).

A Gaussian Process is a collection of random variables, any finite subset of which has a joint Gaussian distribution. Gaussian Processes are fully parametrised by a mean function and covariance function, or kernel. Given a real process $f(\mathbf{x})$, we write the Gaussian Process as:

$$f(\mathbf{x}) \sim \mathcal{GP}(m(\mathbf{x}), k(\mathbf{x}, \mathbf{x'})) \quad (7)$$

where functions $m(\mathbf{x})$ and $k(\mathbf{x},\mathbf{x'})$ are respectively the mean and covariance functions:

$$m(\mathbf{x}) = \mathbb{E}[f(\mathbf{x})] \quad (8)$$
$$k(\mathbf{x}, \mathbf{x}') = \mathbb{E}[(f(\mathbf{x}) - m(\mathbf{x})) \times (f(\mathbf{x'}) - m(\mathbf{x'}))] \quad (9)$$

Inputs are commonly centered during pre-processing. For a given training set $\mathbf{X} = \{\mathbf{x_1}, ..., \mathbf{x_n}\}$ with corresponding output variables $\mathbf{y} = \{y_1, ..., y_n\}^\top$ and Gaussian Process $f$, the distribution of $\mathbf{f} = [f(\mathbf{x_1}), ..., f(\mathbf{x_n})]^\top$ will be multivariate Gaussian:

$$\mathbf{f} \sim \mathcal{N}(\mathbf{0}, \mathbf{K}) \quad (10)$$

where $K_{ij} = k(\mathbf{x_i}, \mathbf{x_j})$. Conditional on $\mathbf{f}$, we have a Gaussian observation model given by:

$$y_i | f(\mathbf{x_i}) \sim \mathcal{N}(\mathbf{0}, \sigma_n^2) \quad (11)$$

where $\sigma_n^2$ parametrises noise. Gaussian distribution conjugacy allows us to marginalise out $\mathbf{f}$ to find the distribution:

$$y_i \sim \mathcal{N}(\mathbf{0}, \mathbf{K} + \sigma_n^2 \mathbf{I}) \quad (12)$$

and conditioning on the training data yields the following predictive distribution $y^*$ for an unseen test datapoint $\mathbf{x}^*$:

$$y^* | \mathbf{x}^*, \mathbf{X}, \mathbf{y} \sim \mathcal{N}(\mathbf{k}^*(\mathbf{K} + \sigma_n^2 \mathbf{I})^{-1} \mathbf{y},$$
$$k^{**} - \mathbf{k}^*(\mathbf{K} + \sigma_n^2 \mathbf{I})^{-1} \mathbf{k}^{*\top}) \quad (13)$$

where $K_{ij} = k(\mathbf{x_i}, \mathbf{x_j})$, $\mathbf{k}^* = [k(\mathbf{x_1}, \mathbf{x}^*), ..., k(\mathbf{x_n}, \mathbf{x}^*)]$ and $k^{**} = k(\mathbf{x}^*, \mathbf{x}^*)$. This methodology combines prior knowledge over $\mathbf{f}$, encoded in the covariance function $k(\mathbf{x}, \mathbf{x}')$, with observation data to produce a posterior distribution for forecasting.

To counter overfitting, we introduce *k*-fold cross-validation, a model validation methodology that involves partitioning the original training set into *k* complementary subsets. We then train the model on *k-1* subsets and test it on the one remaining subset. After rotating through the *k* choices for this validation set, the results are averaged across all tests and provide insight into the model's ability to generalise well. We apply 10-fold cross-validation to determine the optimal covariance function for our dataset from a range of options (Squared Exponential, Rational Quadratic, Matérn 1/2, Matérn 3/2 and Matérn 5/2; Rasmussen and Williams, 2006), and settle on the Matérn 3/2 kernel, a once-differentiable function exhibiting the low smoothness typical of financial time series.

$$k(\mathbf{x}, \mathbf{x}') = \sigma_f^2 \left(1 + \frac{\sqrt{3}|\mathbf{x} - \mathbf{x}'|}{l}\right)$$
$$\times \exp\left(-\frac{\sqrt{3}|\mathbf{x} - \mathbf{x}'|}{l}\right) \quad (14)$$

The covariance function above employs an isotropic Manhattan norm as the similarity measure between two vectors in input space. This assumes that a single, global characteristic length scale *l* can appropriately evaluate proximity in all input dimensions. Even with all inputs normalised to the same scale during pre-processing, it is likely that they will contain varying levels of information on the output variable, motivating the use of input-specific characteristic length scales.

In ARD kernels, the scalar input length scale $l$ of Equation (14) is replaced with a vector input length scale with a different $l_i$ for each input dimension $i$, allowing for different distance measures. These hyperparameters will adapt to any given dataset: inputs with large length scales cause only marginal variations in the covariance function, whereas inputs with small length scales effectively magnify those variations. We can therefore define the relevance score of each feature to be the reciprocal of its input length scale, and rank the salience of inputs by descending relevance.

$$\text{Relevance Score}_i = l_i^{-1} \qquad (15)$$

ARD algorithms have been successfully used in research ranging from bioinformatics (Campbell and Tipping, 2002) to seismography (Oh et al., 2008), providing an effective tool for pruning large numbers of irrelevant features. A limitation of the methodology as presented is that the relevance scores only provide a relative ranking between the features of a model. Two equally meaningless inputs will have relevance scores of similar magnitude, as would two equally meaningful features. On their own, these scores provide little basis for performing dimensionality reduction. To overcome this, we include in each regression a baseline feature composed of standard Gaussian noise. We assert that a meaningful input should have a relevance score that is at least two orders of magnitude greater than noise, so by computing the Relevance Ratio we can determine which features are objectively informative.

$$\text{Relevance Ratio}_i = \frac{\text{Relevance Score}_i}{\text{Relevance Score}_{noise}} \qquad (16)$$

## 5. Results

In this section we outline the findings of our analysis. We begin by discovering relevance hierarchies in the data using ARD, before proceeding with model testing and benchmarking. Model performance metrics were derived using market data from Jan-13 to Dec-14 for training and Jan-15 to Apr-15 for testing. The price history of the S&P500 Index for this period is provided in Fig. 1.

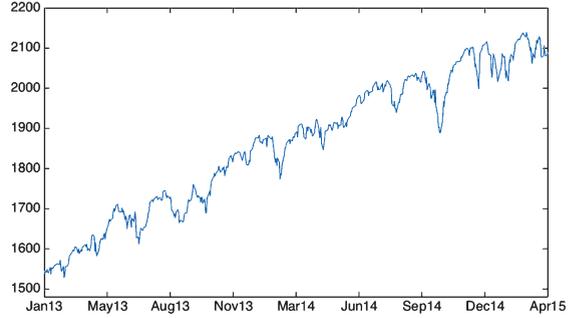

Fig. 1. S&P500 Index price history between Jan-13 and Apr-15.

### 5.1. Correlation analysis

We begin by running a correlation analysis on each feature of the training set, grouped by domain and collect the findings in Table 1. In most cases, rank correlations are stronger than linear correlations, though the variations are too marginal to alter the analysis. For brevity, in all ensuing sections we have adopted the linear definition of correlation.

We next outline a methodology for determining whether an observed sample correlation is significant. Given two independent random variables $x_i$ and $y$ of length $N$ with sample correlation $r$, the statistic

$$t = \frac{r \times \sqrt{N-2}}{\sqrt{1-r^2}} \qquad (17)$$

is t-distributed with *N-2* degrees of freedom. Values for the *(r, N)* pair that land outside the 95% confidence interval of the t-distribution violate the null hypothesis of independence, providing a methodology via the Student's *t*-test for identifying significant correlations in a dataset. *P*-values are derived from t-distribution tables and measure the probability that uncorrelated sample data will yield a t-statistic as or more extreme than the value of *t* obtained from Equation (17). Common significance thresholds are *p*-values of 0.05 or 0.01. Applying *t*-tests to our dataset, every domain apart from broker recommendations held at least one feature bearing significant correlation with next-day returns, signalled by p-values under 0.05 in Table 1.

The use of 4 distinct domains stemmed from the belief that, by virtue of tracking different market agents, these datasets will exhibit low correlation with each other and therefore enhance the predictive power of a combined model. In Table 2 we measure the correlation between input pairs in the training set, and find indeed that *intra-domain* correlations are generally stronger than *inter-domain*

Table 1
Input-Output correlation analysis measured on the training set, $N = 503$ (Jan-13 to Dec-14)

| Feature | Correlation | | $p$-value | |
| --- | --- | --- | --- | --- |
| | Pearson | Spearman | Pearson | Spearman |
| Return(t) | −0.0336 | −0.0862 | 0.4524 | 0.0534 |
| **50dSMA** | **−0.0451** | **−0.1123** | **0.3130** | **0.0117** |
| **MACD** | **−0.1403** | **−0.1576** | **0.0016** | **0.0004** |
| Signal Line | −0.0170 | −0.0365 | 0.7034 | 0.4138 |
| **Stocktwits** | **−0.1103** | **−0.1247** | **0.0133** | **0.0051** |
| Twitter | −0.0287 | −0.0539 | 0.5201 | 0.2275 |
| Stocktwits Change | −0.0581 | −0.0658 | 0.1933 | 0.1406 |
| Twitter Change | +0.0269 | +0.0215 | 0.5474 | 0.6305 |
| **Directionality** | **+0.1011** | **+0.1135** | **0.0234** | **0.0108** |
| **Viscosity*** | **−0.2262** | **−0.1831** | **0.0001** | **0.0001** |
| Broker State | +0.0348 | +0.0159 | 0.4361 | 0.7220 |
| Broker Change | +0.0024 | +0.0263 | 0.9564 | 0.5562 |

*Correlation for Viscosity was calculated against absolute returns.

Table 2
Input-Input correlation analysis measured on the training set, $N = 503$ (Jan-13 to Dec-14)

| | Technicals | | | | Sentiment | | Price Space | | Broker | |
| --- | --- | --- | --- | --- | --- | --- | --- | --- | --- | --- |
| | yret | 50dMA | MACD | Signal | Twtr | ST | Dir | Visc | State | Change |
| yret | 1.00 | 0.34 | −0.03 | 0.18 | 0.29 | 0.29 | 0.02 | 0.01 | 0.05 | 0.01 |
| 50dMA | 0.34 | 1.00 | 0.13 | 0.08 | 0.11 | 0.12 | −0.11 | 0.02 | 0.05 | 0.00 |
| MACD | −0.03 | 0.13 | 1.00 | 0.49 | 0.11 | 0.24 | −0.49 | 0.37 | −0.01 | −0.15 |
| Signal | 0.18 | 0.08 | 0.49 | 1.00 | 0.27 | 0.27 | −0.18 | 0.18 | 0.10 | −0.12 |
| Twtr | 0.29 | 0.11 | 0.11 | 0.27 | 1.00 | 0.44 | −0.14 | 0.21 | 0.15 | −0.01 |
| ST | 0.29 | 0.12 | 0.24 | 0.27 | 0.44 | 1.00 | −0.18 | 0.11 | 0.05 | 0.02 |
| Dir | 0.02 | −0.11 | −0.49 | −0.18 | −0.14 | −0.18 | 1.00 | −0.35 | −0.07 | 0.03 |
| Visc | 0.01 | 0.02 | 0.37 | 0.18 | 0.21 | 0.11 | −0.35 | 1.00 | 0.02 | −0.10 |
| State | 0.05 | 0.05 | −0.01 | 0.10 | 0.15 | 0.05 | −0.07 | 0.02 | 1.00 | 0.11 |
| Change | 0.01 | 0.00 | −0.15 | −0.12 | −0.01 | 0.02 | 0.03 | −0.10 | 0.11 | 1.00 |

correlations, inspiring the pursuit of information gain across diverse, heterogeneous datasets.

### 5.2. Feature relevance

Using training data from Jan-13 to Dec-14, we implement separate Gaussian Process regressions for each data domain using the Matérn 3/2 ARD kernel. This allows both a ranking of feature relevance within each domain, and bivariate visualisations of the mean surfaces learned from the two top-ranked features in each model. Relevance is ranked on the basis of Relevance Score and Relevance Ratio defined in Equations (15) and (16) respectively, with results for market technicals provided in Table 3.

Whilst the MACD-derived signal line and previous day's return explained little of the variation in output, the 50-day Simple Moving Average was salient, as was the MACD. Figure 2 provides a heatmap of return variation based on the two top features of the technical domain, MACD and 50dMA(t), indexed by percentile score. As a first approximation, MACD and next-day returns move inversely: cheapness with respect to recent history correlates with next-day gains.

Table 3
Relevance of market technicals

| | Relevance | |
| --- | --- | --- |
| Feature | Score | Ratio |
| Return(t) | 0.0637 | $4.3 \times 10^2$ |
| 50dSMA | 0.5620 | $3.8 \times 10^3$ |
| MACD | 0.1783 | $1.2 \times 10^3$ |
| Signal Line | 0.0883 | $6.0 \times 10^2$ |
| Noise | 0.0002 | 1 |

Table 4 provides an analysis of sentiment feature relevance. Stocktwits sentiment data is significantly more informative than Twitter data, to the point where the Twitter feature is irrelevant and can be discarded. As a social media site focused on finance, it is likely that Stocktwits's polarity reflects solely market sentiment, whereas Twitter's captures public opinion on a wide range of market-irrelevant issues (celebrity gos-

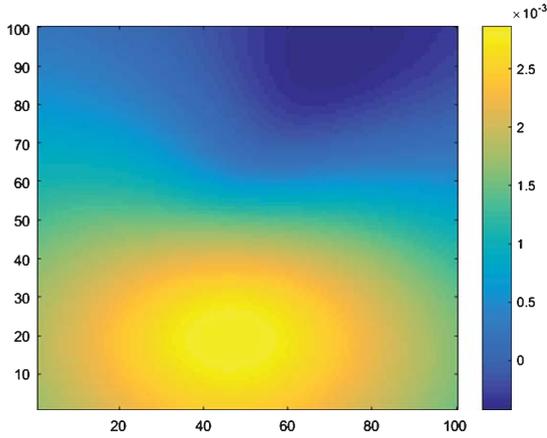

Fig. 2. S&P500 Daily Return Variation as a function of 50-day Moving Average (x-axis) and MACD (y-axis).

Table 4
Relevance of sentiment analysis

| Feature | Relevance Score | Ratio |
| --- | --- | --- |
| Stocktwits Index | 0.2087 | $2, 8 \times 10^3$ |
| Stocktwits Change | 0.0001 | 0.9 |
| Twitter Index | 0.0001 | 0.9 |
| Twitter Change | < 0.0001 | 0.3 |
| Noise | 0.0001 | 1 |

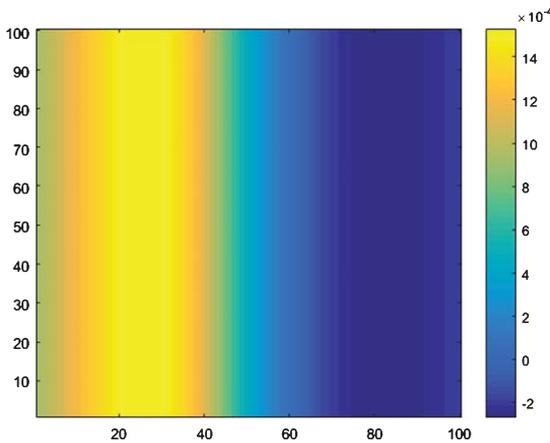

Fig. 3. S&P500 Daily Return Variation as a function of Stocktwits Sentiment (x-axis) and Twitter Sentiment (y-axis).

sip, local politics). The 1-day change variables were also meaningless and can be discarded from subsequent analysis. Notably, the mean function learned through GP regression calls into question the wisdom of crowds: as Fig. 3 indicates, optimism on Stocktwits foreshadows broad market declines, and conversely. Sentiment analysis lends credence to the

Table 5
Relevance of price space

| Feature | Relevance Score | Ratio |
| --- | --- | --- |
| Directionality | 0.5656 | $4.7 \times 10^3$ |
| Viscosity | 0.3844 | $3.2 \times 10^3$ |
| Noise | 0.0001 | 1 |

Warren Buffett adage: "be greedy when others are fearful, fearful when others are greedy."

Relevance for options-derived metrics is provided in Table 5. Directionality and viscosity were almost equally relevant, with positive directionality - that is, experts pre-positioning for rallies via call options - anticipating positive next-day returns. Viscosity instead tracked areas of return compression, and acted as a form of friction. This manifests in Fig. 4 as areas of peak return coinciding with high directionality and low viscosity.

The relevance of broker actions is assessed in Table 6. Broker upgrades and downgrades are infrequent occurrences, resulting in a sparse Broker Change input. The Matérn 3/2 kernel is capable of learning the non-smooth behaviour exhibited in Fig. 5, but with relevance metrics indistinguishable from Gaussian noise, it is unlikely this domain will provide meaningful improvements to a combined model. This suggests that analyst opinions have little predictive power, and merely reflect market changes after they've occurred.

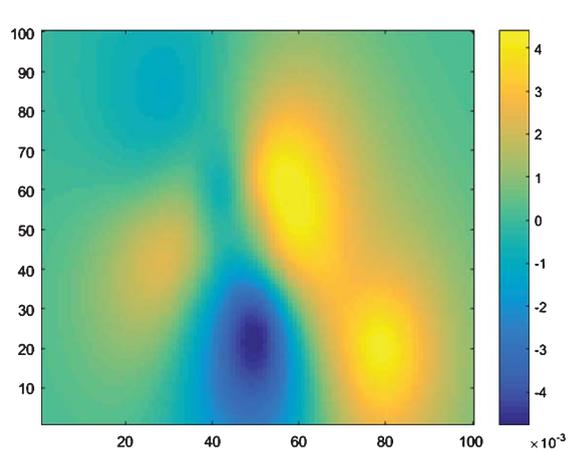

Fig. 4. S&P500 Daily Return Variation as a function of Directionality (x-axis) and Viscosity (y-axis).

Table 6
Relevance of broker recommendations

| Feature | Relevance Score | Ratio |
| --- | --- | --- |
| Broker State | 0.0157 | $2.0 \times 10^{-2}$ |
| Broker Change | 0.2649 | $0.3 \times 10^{-1}$ |
| Noise | 0.4523 | 1 |

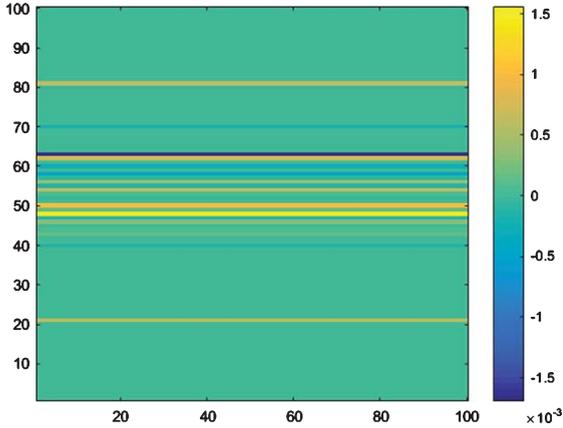

Fig. 5. S&P500 Daily Return Variation as a function of Broker State (x-axis) and Broker Change (y-axis).

Retaining only the salient features, we run a high-dimensional Gaussian Process regression on relevant inputs from all domains simultaneously. The results, compiled in Table 7, broadly mirror our expectations from the correlation analysis, highlighting the ARD framework's ability to discover structure in the data.

### 5.3. Model performance

Having established a method for identifying salient features, we now turn our attention to the predictive performance of ARD Gaussian Processes using each data domain. We separately test the predictive value of each domain before fusing them into a combined model, and measure performance according to the Pearson correlation between forecasts and observations, Median Absolute Deviation and Normalised Root Mean Square Error, where the normalisation constant is the standard deviation of the observations. The results are provided in Table 8.

The model registers monotonic improvements in performance when additional features are included, with the options market data providing the greatest gain. Moreover, it strictly outperforms traditional financial models such as look-ahead AR Processes on measures of ground-truth correlation and NRMSE. Benchmark performance levels are included in Table 9.

Over timeframes much larger than our study's 28-month window, supervised batch algorithms in finance run the risk of failing to recognise significant changes in the landscape fast enough. For example, Stocktwits sentiment's relevance would have been

Table 7
Relevance across all Domains measured on the training set, $N = 503$ entries (Jan-13 to Dec-14)

| | Relevance | | Pearson | |
|---|---|---|---|---|
| Feature | Score | Ratio | Correlation | $p$-value |
| Directionality | 0.3698 | $7.5 \times 10^3$ | +0.1011 | 0.0234 |
| Viscosity* | 0.3332 | $6.7 \times 10^3$ | −0.2262 | 0.0001 |
| Stocktwits | 0.0738 | $1.5 \times 10^3$ | −0.1103 | 0.0133 |
| 50dMA | 0.6660 | $1.3 \times 10^4$ | −0.0451 | 0.3130 |
| MACD | 0.3159 | $6.4 \times 10^3$ | −0.1403 | 0.0016 |
| Broker Change | < 0.0001 | 1.58 | +0.0024 | 0.9564 |
| Noise | < 0.0001 | 1 | −0.0238 | 0.5948 |

*Correlation for Viscosity was calculated against absolute returns.

Table 8
ARD GP Performance measured on the test set, $N = 75$ (Jan-15 to Apr-15)

| | Pearson | | Performance | |
|---|---|---|---|---|
| Feature | Correlation | $p$-value | MAD (bp) | NRMSE |
| **MACD** | **+0.2387** | **0.0392** | **58.22** | **0.9834** |
| Stocktwits | +0.1779 | 0.1268 | 52.51 | 0.9888 |
| **Directionality** | **+0.2412** | **0.0371** | **54.07** | **0.9769** |
| Viscosity* | −0.1635 | 0.1611 | 51.59 | 0.9880 |
| Broker Change | −0.1206 | 0.3026 | 51.73 | 0.9941 |
| **Technicals (all)** | **+0.3079** | **0.0072** | **56.99** | **0.9796** |
| Sentiment (all) | +0.1779 | 0.1268 | 52.51 | 0.9888 |
| **Price Space (all)** | **+0.3315** | **0.0037** | **60.76** | **0.9477** |
| Broker Data (all) | −0.1343 | 0.2505 | 51.73 | 0.9941 |
| **Combined** | **+0.3803** | **0.0008** | **51.53** | **0.9298** |

*Correlation for Viscosity was calculated against absolute returns.

Table 9
Look-ahead benchmark performance

| Model | Correlation | MAD (bp) | NRMSE |
|---|---|---|---|
| AR(1) | +0.0050 | 53.10 | 0.9950 |
| AR(3) | +0.2025 | 53.11 | 0.9932 |
| AR(10) | +0.1950 | 52.61 | 0.9885 |

Table 11
Adaptive benchmark performance

| Model | Correlation | MAD (bp) | NRMSE |
|---|---|---|---|
| AR(1) | +0.1163 | 48.01 | 0.9891 |
| AR(3) | +0.1095 | 49.27 | 0.9887 |
| AR(10) | +0.3191 | 51.70 | 0.9561 |
| KF(1) | +0.0973 | 51.33 | 0.9909 |
| KF(3) | +0.0219 | 49.03 | 0.9940 |
| KF(10) | +0.1952 | 52.74 | 0.9763 |

very low when the site was launched in 2009, and grew in tandem with the size of its user base. A solution to this challenge involves adaptively learning the kernel hyperparameters from recent history only, removing the impact of old, potentially irrelevant data. The evolution from offline to online, adaptive learning follows straightforwardly: we define a window $w$ over which to train an adaptive ARD Gaussian Process for next-day predictions. Rolling the window forward, we generate forecasts for each day in our test set using the optimally combined feature set, and measure model performance as before. Performance metrics for the Adaptive ARD Gaussian Process model are included in Table 10.

Predictive performance dips to impractical levels below the $w = 250$ threshold corresponding to one full year's data, highlighting the need for a critical mass of data for Gaussian Process regression and hinting at seasonal variance in stock market returns, in line with a long history of empirical studies on the topic of annual cyclicality (Lakonishok and Smidt, 1989; Agrawal and Tandon, 1994). Factoring in correlation and Mean Absolute Deviation measures, the best adaptive performance was obtained using exactly one full year of the most recent data.

In Table 11 we provide performance metrics on benchmark adaptive models such as one-step-ahead AR and autoregressive Kalman Filters with varying lags, and find the Adaptive ARD GP yields both superior results and the benefit of automatic, interpretable feature selection.

## 6. Conclusions

Extracting information from multiple domains presents the dual challenge of identifying both what to pick and how to mix. Our results provide a principled framework for reducing input dimensionality through iterative ARD GP regression. We show measurable gains in predictive performance from fusing multiple data streams together in an online setting, and draw particular attention to the relevance of options market data and the implicitly inhomogeneous representation of price space. As an untapped, feature-rich, strike-dependent dataset shaped by the interactions of informed players, options market salience provides a strong mandate for further research into data-driven modelling of price space and its implications for financial forecasting.

Table 10
Adaptive ARD GP performance measured on the test set, $N = 75$ (Jan-15 to Apr-15)

| | Pearson | | Performance | |
|---|---|---|---|---|
| Window Length | Correlation | *p*-value | MAD (bp) | NRMSE |
| $w = 150$ | +0.2922 | 0.0110 | 50.96 | 0.9990 |
| $w = 175$ | +0.3181 | 0.0054 | 44.12 | 0.9769 |
| $w = 200$ | +0.3019 | 0.0085 | 49.08 | 0.9756 |
| $w = 225$ | +0.3147 | 0.0060 | 53.29 | 0.9692 |
| **w = 250** | **+0.3797** | **0.0007** | **43.33** | **0.9377** |
| $w = 275$ | +0.3551 | 0.0018 | 48.64 | 0.9579 |
| $w = 300$ | +0.3368 | 0.0031 | 52.06 | 0.9686 |
| $w = 325$ | +0.2966 | 0.0098 | 61.31 | 0.9789 |
| $w = 350$ | +0.3111 | 0.0066 | 61.62 | 0.9620 |
| $w = 375$ | +0.3236 | 0.0046 | 57.80 | 0.9438 |
| $w = 400$ | +0.3526 | 0.0019 | 54.84 | 0.9313 |
| $w = 425$ | +0.3359 | 0.0032 | 63.02 | 0.9369 |
| $w = 450$ | +0.3584 | 0.0016 | 58.86 | 0.9286 |
| $w = 475$ | +0.3508 | 0.0020 | 57.77 | 0.9313 |
| $w = 500$ | +0.3636 | 0.0013 | 57.93 | 0.9275 |